\documentclass[%
 reprint,
 showkeys,
 showpacs,
 superscriptaddress,
 nofootinbib,
 amsmath,amssymb,
 aps,
]{revtex4-1}

\usepackage[utf8]{inputenc}
\usepackage[english]{babel}
\usepackage{amsmath}
\usepackage{amsfonts}
\usepackage{amssymb}
\usepackage{graphicx}

\newcommand{\rot}{{\mathrm{rot}} \, } 
\newcommand{\derivep}[2]{ \frac{\partial #1}{\partial #2} }

\newcommand{\bfE}{{\mathbf{E}}}

\newcommand{\bfH}{{\mathbf{H}}}
\newcommand{\bfJ}{{\mathbf{J}}}

\newcommand{\bfe}{{\mathbf{e}}}

\newcommand{\bfr}{{\mathbf{r}}}

\newcommand{\rmc}{{\mathrm c}}
\newcommand{\rmb}{{\mathrm b}}

\newcommand{\upd}{{\, \mathrm d}}
\newcommand{\rme}{{\mathrm e}}
\newcommand{\upe}{{\, \mathrm e}}
\newcommand{\rmg}{{\mathrm g}}
\newcommand{\rmi}{{\mathrm i}}

\newcommand{\rmpp}{{\mathrm p}}

\newcommand{\cC}{{\mathcal C}}

\newcommand{\cP}{{\mathcal P}}

\newcommand{\cS}{{\mathcal S}}
\newcommand{\cV}{{\mathcal V}}

\newcommand{\NN}{{\mathbb N}}

\newcommand{\ZZ}{{\mathbb Z}}

\newcommand{\sE}{{\mathsf E}}
\newcommand{\bsE}{{\boldsymbol{\mathsf{E}}}}
\newcommand{\sF}{{\mathsf F}}

\newcommand{\bsH}{{\boldsymbol{\mathsf{H}}}}
\newcommand{\sI}{{\mathsf I}}

\newcommand{\sK}{{\mathsf K}}

\newcommand{\sV}{{\mathsf V}}

\begin{document}
\title{Comparison Between Pierce Equivalent Circuit and Recent Discrete Model for Traveling-Wave Tubes}


\author{Damien~F.~G.~Minenna}%
 \email[Electronic address: ]{damien.minenna@univ-amu.fr}
\affiliation{%
Centre National d'\'Etudes Spatiales, 31401 Toulouse cedex 9, France
}%
\affiliation{%
Aix-Marseille University, UMR 7345 CNRS PIIM, \'equipe turbulence plasma, case 322 campus Saint J\'er\^ome,  av.\ esc.\ Normandie-Niemen, 13397 Marseille cedex 20, France
}%
\affiliation{%
Thales Electron Devices, rue Lat\'eco\`ere, 2, 78140 V\'elizy, France
}%
\author{Artem~G.~Terentyuk}
\altaffiliation[Benefited from]{ an Ostrogradski fellowship from the French embassy in Russia.}%
 \altaffiliation[Acknowledge support from]{ the Russian Science Foundation grant No 17-12-01160.}
\affiliation{%
Saratov State University, 410012 Saratov, Russia, and with Saratov Branch of the Institute of Radio Engineering and Electronics, Russian Academy of Sciences, 410019 Saratov, Russia
}
\author{Fr{\'e}d{\'e}ric~Andr{\'e}}%
 \email[Electronic address: ]{frederic.andre@thalesgroup.com}
\affiliation{%
Thales Electron Devices, rue Lat\'eco\`ere, 2, 78140 V\'elizy, France
}%
\author{Yves~Elskens}%
 \email[Electronic address: ]{yves.elskens@univ-amu.fr}
\affiliation{%
Aix-Marseille University, UMR 7345 CNRS PIIM, \'equipe turbulence plasma, case 322 campus Saint J\'er\^ome,  av.\ esc.\ Normandie-Niemen, 13397 Marseille cedex 20, France
}%
\author{Nikita~M.~Ryskin}%
 \email[Electronic address: ]{ryskinnm@info.sgu.ru} 
 \altaffiliation[Acknowledge support from]{ the Russian Science Foundation grant No 17-12-01160.}
\affiliation{%
Saratov State University, 410012 Saratov, Russia, and with Saratov Branch of the Institute of Radio Engineering and Electronics, Russian Academy of Sciences, 410019 Saratov, Russia
}

\date{October 31, 2017}

\begin{abstract}
To perform accurate numerical simulations of the traveling-wave tube
in time domain, a new approach using field decomposition with large
reduction of degrees-of-freedom has been proposed: the discrete
model. To assess its validity, we compare it with the well-established
Pierce equivalent circuit model in small signal regime. We also
discuss associated beam, circuit-beam, and circuit impedances. We
demonstrate analytically and with a numerical example that the newly
developed discrete model is very close to the Pierce model. 
Interestingly, small deviations do exist at the edges of the
amplification band. We speculate that the deviation from reality is on
the Pierce model side, while the discrete model would be more
accurate.
\end{abstract}

    
\maketitle


\section{Introduction}
\label{Intro}

The recently developed discrete model
(a.k.a.\ Kuznetsov discrete model) \cite{kuz80,rys09,and13} is a
promising tool to analyse devices such as traveling wave tubes (TWTs)
beyond what is possible today with the well established Pierce
model\cite{pie50}.
It provides an exact reduction of degrees-of-freedom for
electromagnetic fields and allows to build both frequency \cite{ter16}
and time domain algorithms \cite{min16} that are faster and more
accurate alternatives to current PIC algorithms \cite{and15}.
The discrete model offers several new features compared to 
Pierce's well-known equivalent circuit model.  Most importantly, it is
originally in time domain and enables simulating broadband telecom
signals for example.  Another example are drive-induced oscillations
where spurious frequencies are generated very far from the drive
frequency in the nonlinear regime. This situation will be accessible
to simulation thanks to this new model.  Second, the complex structure
of stop bands can be accurately described and simulated thus offering
a way to progress on the associated oscillation problems.
Fundamentally, the discrete model addresses (and originates from) the
general situation of periodic, quasi-periodic and chaotic particles
interacting with fields, which is of interest to a broader community
of physicists and engineers. Also, three-dimensional simulations are
possible with the discrete model. We deepen these aspects in the
appendices.
 
Before addressing these more complex situations, the first question is
how the new model compares to the existing one in the simplest case of
a single carrier operation (i.e. in frequency domain) in the linear
regime, the original background of the Pierce theory. This is the
objective of this paper.

In section \ref{s:equiCir}, we revisit fundamental definitions of beam, wave and circuit impedances 
from the Pierce equivalent circuit, starting with a model involving only space
charge fields, and then adding circuit fields.
In section \ref{s:Dimo}, we recall the principles of the discrete model
and apply them in the harmonic domain to obtain associated
impedances. Finally, we compare both models in section
\ref{s:comparECandDIMO}.  Appendix~\ref{s:SheaHelix} revisits the
sheath helix approximation using the discrete
model. Appendix~\ref{s:BeamPlasma} compares the TWT discrete model and
beam-plasma models.

\begin{figure}[!t]
\centering
\includegraphics[width=2.5in]{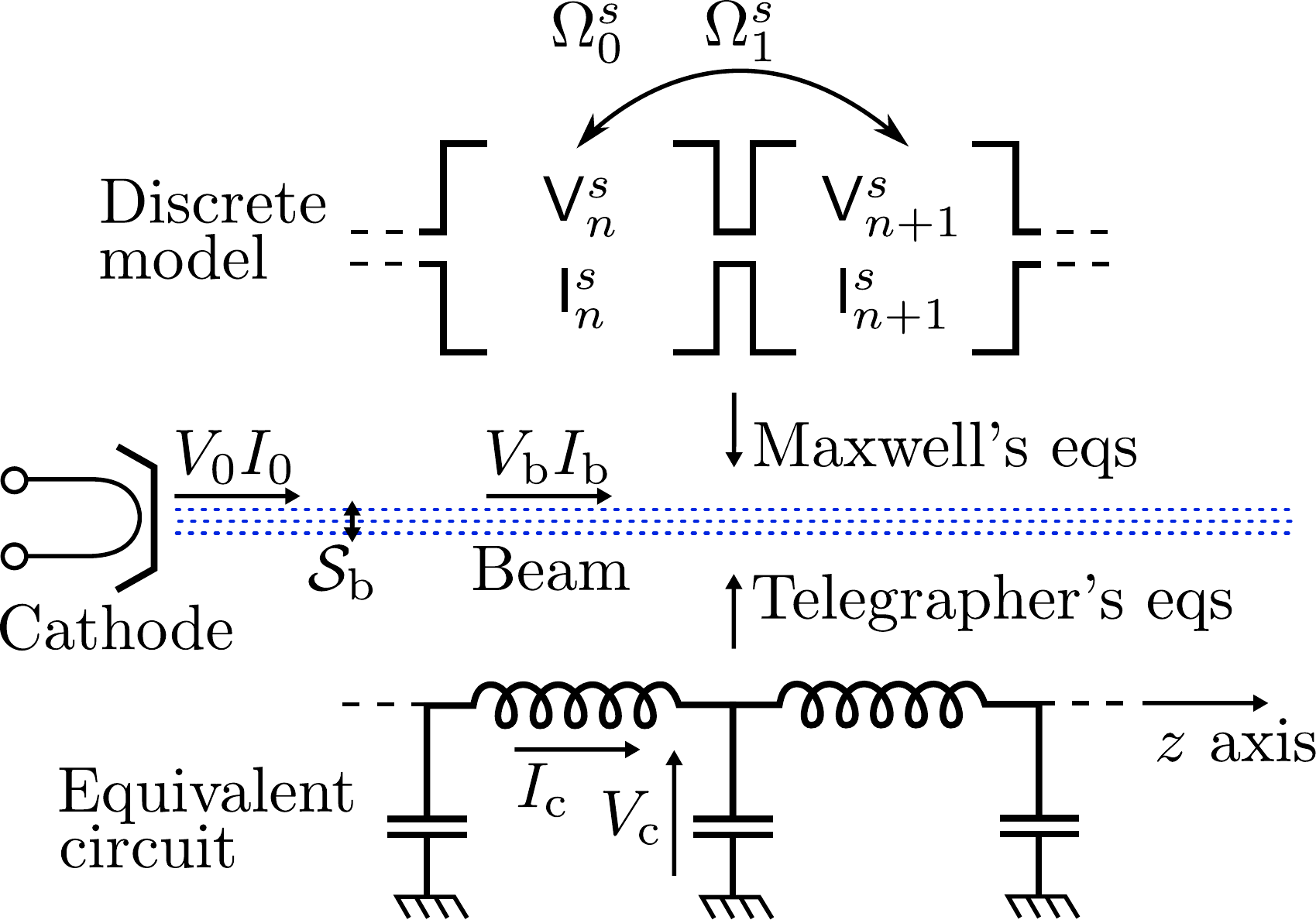}
\caption{Wave-particles interaction for a periodic slow-wave structure represented using the discrete model (above) and the equivalent circuit (below), along the longitudinal $z$-axis. The beam is assumed to be a weakly perturbed fluid with section area $\cS_\rmb$. Variables $V_{\rmc,\rmb,0}$ and $I_{\rmc,\rmb,0}$ are the potential and current of the circuit, beam and cathode (dc beam) respectively. $\sV^s_n$ and $\sI^s_n$ are the temporal variables of electromagnetic circuit fields (see eqs \eqref{e:Efield} and \eqref{e:Hfield}) at cell $n$ for the propagation mode $s$. $\Omega^s_m$ is the coupling coefficient between cells at range $m$.}
\label{f:EquivCir}
\end{figure}

\section{Equivalent circuit} 
\label{s:equiCir}

Developments leading to eqs \eqref{e:Conti} and \eqref{e:Ecircuit}
below are similar to those used in the coupled wave system of
Louisell \cite{lou60} and to the classical analysis by Gilmour \cite{gil94}. 
We reformulate them to facilitate the comparison with the discrete model analysis and keep our paper self-contained. 
In particular, after reaching the dispersion relation
of the Pierce equivalent circuit, we focus on impedances.  
A coupled system composed of a beam (b) and a circuit (c)
will have two electric potentials $V_\rmb$ and $V_\rmc$, and two
different currents $I_\rmb$ and $I_\rmc$, leading a priori to four
different impedances, respectively the beam impedance $Z_{\rmb}$, the
circuit-beam impedance $Z_{\rmc \rmb}$, the beam-circuit impedance
$Z_{\rmb \rmc}$ and the circuit impedance $Z_{\rmc}$.

\subsection{Space charge waves}

The electron beam is described as a weakly perturbed fluid carrying
space charge waves along the longitudinal $z$ coordinate.  Time and
space dependent variables are expressed, according to the space-time
Fourier representation, as $F(z,t) = \Re \left( \tilde F(\beta,\omega)
\upe^{ -\rmi \psi} \right)$, with the local phase $\psi = \beta z
-\omega t$, where $\beta = \omega / v_{\mathrm{ph}}$ is the
propagation constant in the longitudinal direction and $\omega$ the
wave pulsation for the phase velocity $v_{\mathrm{ph}}$. Since we may
study non-resonant regimes, one also defines the electronic
propagation constant $\beta_{\rme} = \omega / v_0$, using the beam
velocity.  Particle velocities are $v_0 + \Re (\tilde v\upe^{ -\rmi \psi})$,
where the initial velocity $v_0 = \sqrt{2 V_0 \eta}$ depends on the
cathode (dc beam) potential $V_0 > 0$ and the charge/mass ratio $\eta
= |e|/m_{\rme}$.  
Particle charge densities are $\rho_0 + \Re (\tilde \rho \upe^{ -\rmi \psi})$, 
with initial density $\rho_0 = I_0 / (v_0 \cS_{\rmb}) < 0 $, 
for a cathode (dc beam) current $I_0 < 0$, and
section area of the beam $\cS_{\rmb}$.  In the linear regime, the
relation between perturbed current density and charge density is
$\tilde J_z = \rho_0 \tilde v + v_0 \tilde \rho$.  As a first step, we
combine this relation with the continuity equation to obtain
\begin{equation} \label{e:Conti}
\left( \omega - \beta v_0 \right) \cS_{\rmb} \tilde J_z = - \omega \,
\frac{|I_0|}{2 V_0} \, \tilde V_{\rmb} \, ,
\end{equation}
with the perturbed beam potential $\tilde V_{\rmb} = v_0 \tilde v /
\eta$.  The minus sign comes from the dc current $I_0 < 0$.  The
continuity equation remains unchanged by the presence of circuit
waves, so we will keep eq.~\eqref{e:Conti} in the next section.

On the other hand, if we only consider space charge waves in our
system (neglecting metallic boundary conditions for simplicity), the
Euler equation for electron motion provides $(\rmi \omega - \rmi \beta
v_0 )\tilde v = - \eta \tilde E_{z, \mathrm{sc}}$, with the space
charge field $\tilde E_{z, \mathrm{sc}} = \rmi \tilde J_z /
(\epsilon_0 \omega) $ from Poisson and continuity
equations. Therefore, this motion equation is rewritten
\begin{equation} \label{e:MotionEqSC}
 \left(\omega - \beta v_0 \right) \, \frac{\tilde V_{\rmb}}{v_0} =
 \frac{\omega^2_{\rmpp}}{\omega v_0} \, \frac{2 V_0}{|I_0|} \,
 \cS_{\rmb} \tilde J_z \, ,
\end{equation}
with the electron plasma pulsation $\omega_\rmpp = \sqrt{ \eta
  |\rho_0| / \epsilon_0}$. Equation \eqref{e:MotionEqSC} must be equal
to $- \tilde E_{z, \mathrm{sc}}$.  Now, we rewrite the relation
between the space charge field and the electron current as  
\begin{equation} \label{e:E=ibZSJ}
\tilde E_{z, \mathrm{sc}} = -\nabla \tilde V_\mathrm{sc}= \rmi \beta
\tilde V_\mathrm{sc}= - \rmi \beta Z_\rmb \tilde I_\rmb = - \rmi \beta
Z_{\rmb} \int_{\cS_{\rmb}} \tilde J_z \upd x \upd y \, ,
\end{equation}
defining the beam characteric impedance $Z_{\rmb}$, with a minus sign
from the negative charge density.  Comparing eqs \eqref{e:MotionEqSC}
and \eqref{e:E=ibZSJ} immediatly yields\footnote{The minus sign in
  \eqref{e:Conti}, \eqref{e:E=ibZSJ} and \eqref{e:BeamImpedance} come
  from our notation $I_0 < 0$. This result reads $Z_{\rmb} =
  \frac{\omega_\rmpp}{\omega} \, \frac{2 V_0}{| I_0 | }$ in
  ref.~\cite{lou60}, but Louisell was working in the reference frame
  of the beam instead of the laboratory frame as here.}
\begin{equation} \label{e:BeamImpedance}
Z_{\rmb}(\beta) =  \frac{ - \omega_\rmpp^2}{\omega \beta v_0} \, \frac{2 V_0}{| I_0 | }  \, ,
\end{equation}
as $2 V_0/| I_0 | = v_0 / (|\rho_0| \eta \cS_{\rmb})$, and if we
insert \eqref{e:Conti} into \eqref{e:E=ibZSJ}, we have $(\omega -
\beta v_0)^2 = \omega_\rmpp^2$, viz.\ the cold\footnote{In the plasma context, ``cold'' means neglecting the beam temperature (and pressure) in its ballistic co-moving frame.}
Bohm-Gross dispersion
relation \cite{boh49} for space charge waves only.  They are
represented in Fig.~\ref{f:DispRelationNoAmp}. The ratio $V_0 /
I_0$ is the beam impedance in case of unperturbed beam ($\tilde J_z =
\tilde V_{\rmb} = 0$), so we refer to it as the cathode (dc)
impedance.

\subsection{Coupling to slow-wave circuits}

Now, we consider the equivalent circuit model (see
Fig.\ \ref{f:EquivCir}) provided by \cite{pie50} in the small signal
regime, and we add circuit waves to the previous system.  In the
motion equation \eqref{e:MotionEqSC}, we simply add to the right-hand
side the term $-\tilde E_{z,\rmc}$, corresponding to the electric
field from the circuit, and combine eqs \eqref{e:Conti} and
\eqref{e:MotionEqSC} to find 
\begin{equation} \label{e:Ecircuit}
\tilde E_{z,\rmc} = - \rmi \frac{1}{\omega v_0} \, \left[ \left(\omega
  - \beta v_0 \right)^2 - \omega_\rmpp^2 \right] \frac{2 V_0}{| I_0 |
} \, \cS_{\rmb} \tilde J_z \, .
\end{equation}
This is similar to eq.~\eqref{e:E=ibZSJ}, on replacing the space
charge field with the circuit field and the beam impedance with the
circuit-beam impedance $Z_{\rmc \rmb}$ corresponding to the
response of the circuit potential to the beam current, which is
then defined as
\begin{equation}\label{e:CouplingImp}
Z_{\rmc \rmb}(\beta) = \frac{(\omega - \beta v_0)^2 -
  \omega_\rmpp^2}{\omega \beta v_0} \, \frac{2 V_0}{| I_0 |} \, .
\end{equation}
At the resonance, where $\beta_{\rme} = \beta$ (phase velocity equal
to beam velocity), $Z_{\rmc \rmb}$ acts like the beam impedance as if
there were only space charge fields.  Since we have here circuit
waves, we recall the link between eq.~\eqref{e:CouplingImp} and
Pierce's circuit impedance \cite{pie50}
\begin{align}
Z_{\rmc}(\beta) 
&= \frac{|\tilde E_{z,\rmc}|^2}{2 \beta^2 \langle P \rangle}  \label{e:PierceImp} \\
&= \frac{4 V_0}{| I_0 |} \, \cC^3_{\rmpp}  \label{e:PierceImp2} \, ,
\end{align}
with $\cC_{\rmpp}$ the Pierce coupling (or gain) parameter, and
$\langle P \rangle$ the harmonic power. Eq.~\eqref{e:PierceImp} comes
directly from $\tilde V_{\rmc} / \tilde I_{\rmc}$, and it is used by
Pierce to find eq.~\eqref{e:PierceImp2} where the coupling impedance
remains hidden. It would be erroneous to think that for a beamless
case ($V_0 = I_0 = 0$), the circuit impedance could be ill-defined :
following eq.~\eqref{e:PierceImp}, this is not true. In
fact, eq.~\eqref{e:PierceImp2} can only be used for cases with an
existing beam: the Pierce parameter compensates the effect of the
unperturbed beam impedance. This is why the parameter expressing the
coupling of the beam with the circuit is the Pierce coupling
parameter $\cC_{\rmpp}$, not the coupling impedance $Z_{\rmc}$.

\subsection{Telegrapher's equations}

There is another way to find the coupling impedance.  The equivalent
circuit considered on Fig.~\ref{f:EquivCir} is composed of an infinite
number of inductances $L$ and capacitances $C$ per unit length, giving
the evolution equations of the circuit potential and current from
lossless telegrapher's equations (coupled to the beam current)
\begin{align}
- \rmi \beta\tilde V_{\rmc} &= -  \rmi L \omega \tilde I_{\rmc} \, , \\
- \rmi \beta \tilde I_{\rmc} &= - \rmi C \omega \tilde V_{\rmc} + \rmi \beta \tilde I_{\rmb} \, .
\end{align}
Without beam ($\tilde I_{\rmb}=0$),
the uncoupled circuit propagation constant is $\beta_0 = \omega \sqrt{C L}$, and we
find $L \omega = Z_{\rmc} \beta_0$ when recalling the classical
definition of the characteristic impedance $Z_{\rmc} = \tilde V_{\rmc} / \tilde I_{\rmc} = \sqrt{L/C}$
which Pierces defines as the circuit impedance.  Then we merge
the two telegrapher's equations and write the circuit-beam impedance
$Z_{\rmc \rmb} = \tilde V_{\rmc} / \tilde I_{\rmb}$, to find
\begin{equation} \label{e:CouplingImpEvoCir}
Z_{\rmc \rmb} (\beta) = \frac{\beta_0 \beta}{\beta^2_0  -  \beta^2 } Z_{\rmc} \, ,
\end{equation}
equal to  eq.\ \eqref{e:CouplingImp}.
On combining eqs \eqref{e:CouplingImp} and \eqref{e:CouplingImpEvoCir}
with definition \eqref{e:PierceImp2}, we obtain the ``hot" linear
dispersion relation
\begin{equation}\label{e:C3PierceCirq}
\cC^3_{\rmpp} = \frac{(\beta_{\rme} - \beta)^2 - \beta_\rmpp^2}{2
  \beta_{\rme} \beta} \, \frac{\beta^2_0 - \beta^2 } {\beta_0 \beta}
\, ,
\end{equation}
as defined (but written differently) in \cite{pie50}, with
$\beta_{\rme} = \omega / v_0$ and $\beta_\rmpp = \omega_\rmpp/ v_0$.
Equation \eqref{e:C3PierceCirq} exhibits the product of two fractions:
one originating from the beam, and the other one from the circuit.  
It is of the fourth degree, yielding the four natural modes of
propagation. For later use, we rewrite it as
\begin{equation}\label{e:C3PierceCirq2}
\cC^3_{\rmpp} = \frac{(\omega - \beta v_0)^2 - \omega_\rmpp^2}{2
  \omega \beta v_0} \, \frac{\omega^2 - \beta^2 v^2_{\mathrm{ph},0}}
   {\omega \beta v_{\mathrm{ph},0}} \, ,
\end{equation}
with the beamless phase velocity $v_{\mathrm{ph},0} = 1 / \sqrt{C L}$.

\section{Discrete model} 
\label{s:Dimo}

\subsection{Time domain discrete model}

In this section, we briefly revisit basic equations of the Kuznetsov
nonlinear discrete theory \cite{kuz80,rys09,and13}. In the most
general case of any time dependent circuit fields $\bfE(\bfr,t)$,
$\bfH(\bfr,t)$ existing in the delay line (e.g. propagating or
evanescent), we are searching an \emph{exact} and \emph{discretized}
decomposition of that field. To do so, we proceed in three steps. The first
step is that we already know some particular waves propagating in the
structure in the form of the propagation modes. The propagation modes
are calculated as the eigenvectors of the Helmholtz equation with the
Floquet condition at both ends of one period of the
structure\footnote{Propagating modes can be computed thanks to general
  purpose electromagnetic solvers like CST microwave studio or
  HFSS.}. 
The complex envelopes of the propagation mode are written
$\bsE^s_{\beta}(\bfr)$ and $\bsH^s_{\beta}(\bfr)$ where $\beta d$ is
the phase-shift in the Floquet condition and $s \in \NN$ is the label
of the mode. Eigenfields $\bsE^s_{\beta}$ and $\bsH^s_{\beta}$
satisfy the normalization\footnote{In \cite{and13}, this normalisation
  is chosen equal to the eigenfield pulsation $\Omega^s_{\beta}$ so
  that the canonical variables of the Hamiltonian (not discussed here) are the field
  coefficients $\sV^{s}_{n}$ and $\sI^{s}_{n}$ in \eqref{e:Efield}-\eqref{e:Hfield}~; 
  their dimension is then the square root of an action. 
  In \cite{rys09}, this normalisation
  has the dimension of an energy, and $\sV^{s}_{n}$ and $\sI^{s}_{n}$
  become dimensionless.} 
\begin{equation} \label{e:NormalisationNsb}
N^s_{\beta} \delta^s_{s'} = \int_{\cV_{0}} \epsilon_0 \bsE^s_{\beta}
\cdot \bsE^{s'*}_{\beta} \upd^3 \bfr = \int_{\cV_{0}} \mu_0
\bsH^s_{\beta} \cdot \bsH^{s'*}_{\beta} \upd^3 \bfr \, ,
\end{equation}
where $\cV_{0}$ is the cell volume, and $\delta^s_{s'}$ is the
Kronecker symbol.

In a second step, we limit our search for the discretized
expansion to the case of fields $\bfE_\beta(\bfr,t)$
satisfying the Floquet condition (for a phase-shift $\beta d$ per
period). The propagation modes are eigenvectors of the Helmholtz
linear system, with eigenvalues $\Omega^s_{\beta}$,
\begin{align}
\rot \bsE^s_{\beta}(\bfr) &= - \mu_0 \Omega^s_{\beta} \bsH^s_{\beta}(\bfr) \, , \label{e:Helmo1} \\
\rot \bsH^s_{\beta}(\bfr) &= \epsilon_0 \Omega^s_{\beta} \bsE^s_{\beta}(\bfr) \, . \label{e:Helmo2}
\end{align}
As the Helmholtz operator is hermitian, they constitute a vector basis
and we write $\sV_\beta^s(t)$ the discretized set of field
generalized coordinates:
\begin{equation}
  \bfE_\beta(\bfr,t)=\sum_s \sV_\beta^s(t) \bsE_\beta^s(\bfr)  \, .
  \label{e:devel}
\end{equation}
This relation is valid in the reference cell $\cV_{0}$ but all
functions satisfy the Floquet condition, so it is valid everywhere.

The problem now is that fields in general do not respect the Floquet
condition. So our third step is to find an expansion of 
arbitrary fields over a set of fields satisfying the Floquet
condition which would write
\begin{equation}
\bfE(\bfr,t)=\int_{\beta d= - \pi}^{\pi} \bfE_\beta(\bfr,t) \upd (\beta d)  \, .
\label{e:TransGelfandInv}
\end{equation}
Since the $\bfE_\beta$ would satisfy the Floquet condition, we can
rewrite the looked after expansion:
\begin{equation}
\bfE(\bfr + n d \bfe_z,t)=\int_{\beta d=-\pi}^{\pi} \bfE_\beta(\bfr,t)
\rme^{-\rmi n \beta d} \upd (\beta d).
\end{equation}
Thus $\bfE(\bfr + n d \bfe_z,t)$ is the $n^{\textrm{th}}$
coefficient of the Fourier series expansion of $\bfE_\beta$ seen as a
function of $\beta$, namely 
\begin{equation}
 \bfE_\beta(\bfr,t)=\sum_{n \in \NN} \bfE(\bfr + n d \bfe_z,t)
 \rme^{\rmi n \beta d}.
 \label{e:TranfGelfand}
\end{equation}
This yields exactly the looked after $\bfE_\beta$ functions which (i)
satisfy the Floquet condition and (ii) on which the field is
expanded (eq.~\eqref{e:TransGelfandInv}). The elegant transform
\eqref{e:TranfGelfand} into functions satisfying the Floquet condition
was introduced by I. Gel'fand \cite{gel50}, and
\eqref{e:TransGelfandInv} is its inverse transform. It is based on
Fourier series and shares many of its properties. In particular, the
transform of a product is the convolution of the transforms of its
factors. Applying this property to eq.~\eqref{e:devel} completes our
initial search for a discrete model:
\begin{equation}
\bfE(\bfr,t) = \sum_{s \in \NN} \sum_{n \in \ZZ} \sV^{s}_{n}(t) \bsE^s_{-n}(\bfr) \, , \label{e:Efield}
\end{equation}
with $\sV_n^s$ the Gel'fand transform of $\sV_\beta^s$. They are the
discrete variables determining the electric field. The magnetic field
is also discretized\footnote{Ref.~\cite{rys09} uses $\sV^s_{\beta}=
  -\sI^s_{\beta}$ but this is misleading \cite{the16b}. We also use
  $-\pi \leqslant \beta d \leqslant \pi$ instead of $0 \leqslant \beta
  d \leqslant 2 \pi$.}
with its own coordinates $\sI_n^s$
\begin{equation}
\bfH(\bfr,t) = \rmi \sum_{s \in \NN} \sum_{n \in \ZZ} \sI^{s}_{n}(t) \bsH^s_{-n}(\bfr) \, . \label{e:Hfield}
\end{equation}
Note the $\rmi$ factor needed to have real $\sI_n^s$ variables instead
of purely imaginary one.

The interest of this decomposition appears in eqs \eqref{e:Efield} and
\eqref{e:Hfield}. For a single propagating mode, there are $2
n_{\mathrm{max}}$ different time variables (a.k.a.\ degrees of
freedom) for the fields in a delay-line of $n_{\mathrm{max}}$
periods.  In comparison, finite difference techniques used in
particle-in-cell codes necessitate several millions degrees of freedom
to obtain the same accuracy.

We now introduce the beam. Using Maxwell equations with
sources, the field decompositions \eqref{e:Efield}-\eqref{e:Hfield}, 
and the Helmholtz equations \eqref{e:Helmo1}-\eqref{e:Helmo2}, 
we find the evolution equations \cite{and13}
\begin{align}
- \sum_{s \in \NN} \sI^s_{\beta} \Omega^s_{\beta} \bsE^s_{\beta} &=
\sum_{s \in \NN} \derivep{\sV^{s}_{\beta}}{t} \bsE^s_{\beta}
+\frac{\bfJ_{\beta}}{\epsilon_0} - \derivep{\nabla \phi_{\beta}}{t} \,
, \label{e:DimoEvoEq1} \\ \sum_{s \in \NN} \sV^{s}_{\beta}
\Omega^s_{\beta} \bsH^s_{\beta} &= \sum_{s \in \NN}
\derivep{\sI^{s}_{\beta}}{t} \bsH^s_{\beta} \, , \label{e:DimoEvoEq2}
\end{align}
where $\bfJ(\bfr,t)$ is the 3D charge density and the potential
$\phi(\bfr,t)$ satisfies the Poisson equation $\Delta \phi = - \rho /
\epsilon_0$.

\subsection{Harmonic domain discrete model}

In small signal regime, the discrete model in harmonic domain couples the charge density 
$\bfJ (\bfr, t) = \bfJ_0 + \Re (\tilde \bfJ(\bfr) \rme^{\rmi \omega t})$, 
with temporal variables $\sV^s_{\beta} (t) = \tilde{\sV}^s_{\beta} \rme^{\rmi \omega t}$, 
and $\sI^s_{\beta} (t) = \tilde{\sI}^s_{\beta} \rme^{\rmi \omega t}$. 
From eq.~\eqref{e:DimoEvoEq2} and thanks to the eigenfields orthogonality, 
we have $\tilde{\sI}^s_{\beta} = - \rmi \Omega^s_{\beta} \tilde{\sV}^s_{\beta} / \omega$, 
so the evolution equation \eqref{e:DimoEvoEq1} becomes
\begin{equation}
\sum_{s \in \NN} \frac{(\Omega^s_{\beta})^2 - \omega^2}{\omega}  {\tilde{\sV}^s_{\beta}} \bsE^s_{\beta} (\bfr) 
  = 
\frac{- \rmi}{\epsilon_0} \tilde \bfJ_{\beta} (\bfr)  - \omega \nabla \tilde \phi_\beta (\bfr) \, , 
\label{e:EvolutionV}
\end{equation}
where the space charge term $\nabla \tilde \phi_{\beta}$ will disappear under integration over the cell volume thanks to boundary conditions \cite{and13}.
We dot-multiply eq.~\eqref{e:EvolutionV} by the complex conjugate $\bsE^{s*}_{\beta}$ and integrate over space (viz.\ we project on the mode $(s, \beta)$), to find (for a beam with uniform section and small radius)
\begin{equation}
\frac{(\Omega^s_{\beta})^2 - \omega^2}{\omega}  {\tilde{\sV}^s_{\beta}} 
= 
- \rmi S_\rmb \int_0^{d} \tilde J_{z, \beta} (z) \sF^{s*}_{z,\beta} (z) \upd z \, ,\label{e:EvoValongZ}
\end{equation}
with $\sF^{s*}_{z,\beta} (z) =  \sE^{s*}_{z,\beta} (z)  /  N^s_{\beta}$ related to the vector potential eigenfunction.
Eq.~\eqref{e:EvoValongZ} from Maxwell equations replaces the telegrapher's equations in the discrete model. 
We mainly deal with eigenmodes off resonance, so $\Omega^s_{\beta} \neq \omega$ generally. 

To complete our model, we take the same weakly perturbed electron beam as in section \ref{s:equiCir}, so we end with the same linear equation \eqref{e:Ecircuit} but this time with the circuit field \eqref{e:Efield} and the charge density spatially modulated in $z$ by  $\tilde J_z (z) = \hat J_z \rme^{- \rmi \beta z}$.
Then using the inverse Fourier transform, we obtain
\begin{align}
\tilde J_{z, \beta'} (z) &= \sum_{n \in \ZZ} \tilde J_z (z + n d) \upe^{\rmi n \beta' d} = \sum_{n \in \ZZ} \hat J_z \upe^{- \rmi \beta z + \rmi (\beta' - \beta) n d } \nonumber \\
&= \hat J_z \upe^{- \rmi \beta z} \sum_{p \in (2 \pi / d) \ZZ} \delta \left( \frac{\beta' - \beta - p}{2 \pi} d \right) \, , \label{e:JzDimo}
\end{align}
for any wave number $\beta'$, where $\delta$ is Dirac's distribution.
The same is performed for electric field coefficient, $\tilde \sV^s_{\beta'} = \sum_p \hat \sV^s_{p} \delta (\beta' - \beta - p) \frac{2 \pi}{d}$, and \eqref{e:EvoValongZ} becomes
\begin{equation}
\frac{(\Omega^s_{{\beta+p}})^2 - \omega^2}{\omega}  \hat \sV^s_{p} = - \rmi S_\rmb \hat J_z \int_0^{d} \upe^{- \rmi (\beta+p) z} \sF^{s*}_{z,{\beta+p}} (z) \upd z \, . \label{e:EvoIntFkp}
\end{equation}
In the circuit field \eqref{e:Efield}, the integration on $\beta d$ ranges only over $[-\pi, \pi]$, so the sum reduces to the single term $p = 0$ (viz.\ only one band, $s=0$, matters for the waves).
As the Gel'fand eigenfield must respect $\sE^s_{z,\beta} (z) = \hat \sE^s_{z} \upe^{- \rmi \beta z}$ 
(and using $\sF^{s*}_{z,\beta} (0,0,z) = \hat \sE^{s*}_{z} \upe^{\rmi \beta z} / N^s_{\beta}$), 
we have the perturbed circuit field
\begin{equation} \label{e:EcirDIMO}
\tilde E_{z,\rmc} = \sum_{s \in \NN} \hat \sV^s  \sE^s_{z, \beta} (z) \upe^{\rmi \beta z} = \sum_{s \in \NN} \hat \sV^s \hat \sE^s_{z}\, .
\end{equation}
So we finally reach a new expression for \eqref{e:EvoIntFkp}
\begin{equation} \label{e:EVOdiscretMod}
\frac{(\Omega^s_{\beta})^2 - \omega^2}{\omega}  {\hat \sV^s} = - \rmi S_\rmb \hat J_z  d \frac{\hat E^{s*}_{z}}{N^s_{\beta}} \, ,
\end{equation}
and we rewrite eq.~\eqref{e:Ecircuit} with eqs \eqref{e:JzDimo} and \eqref{e:EcirDIMO}.

On the other hand, we can rewrite the circuit impedance from the discrete model as \cite{min17b}
\begin{equation} \label{e:ImpE2Beta}
Z_{\rmc} (\beta) = \frac{|\sE^{s}_{z,\beta}(r=0)|^2 \, d}{\beta^2 v_{\rmg} N^s_{\beta}} \, ,
\end{equation}
where $v_{\rmg}(s,\beta)$ is the group velocity,
and we can compare eq.~\eqref{e:ImpE2Beta} to the equivalent circuit impedance \eqref{e:PierceImp}.
We remark that this wave impedance tends to infinity at the passband edges where the group velocity vanishes. 
An advantage of \eqref{e:ImpE2Beta} is that it involves only experimentally known cold values, 
providing values for $\sE^{s}_{z,\beta}(r=0)$ (and its values in the $n$-representation) from $Z_{\rmc}$.
Following the definition \eqref{e:E=ibZSJ}, but for the circuit field and the circuit-beam impedance, we rewrite the latter for the discrete model
\begin{equation}\label{e:ZcbDIMO}
Z_{\rmc \rmb} (\beta) = - \frac{\rmi}{\beta} \frac{\hat \sV^s \hat \sE^s_{z}}{S_\rmb \hat J_z} \, ,
\end{equation}
for a beam with uniform section.
We insert this relation in eq.~\eqref{e:EVOdiscretMod} and use eq.~\eqref{e:ImpE2Beta} to find a new expression 
\begin{equation} \label{e:NewZcbDIMO}
Z_{\rmc \rmb} (\beta) =  \frac{\omega \beta v_{\rmg}}{\omega^2 - (\Omega^s_{\beta})^2}  Z_{\rmc} \, ,
\end{equation}
enabling us to compare the equivalent circuit--beam coupling impedance \eqref{e:CouplingImp} with the circuit impedance \eqref{e:PierceImp2}. 
Substituting eqs \eqref{e:CouplingImp} and \eqref{e:NewZcbDIMO} in \eqref{e:PierceImp2}, the ``hot" linear dispersion relation of the discrete model becomes
\begin{equation} \label{e:C3PierceDIMO}
\cC^3_{\mathrm{p}} = \frac{(\omega - \beta v_0)^2 - \omega_\rmpp^2}{2 \omega \beta v_0 } \, \frac{\omega^2 - (\Omega^s_{\beta})^2}{\omega \beta  v_{\rmg}} \, .
\end{equation}

\section{Comparison} 
\label{s:comparECandDIMO}

\begin{figure}[!t]
\centering
\includegraphics[width=\columnwidth]{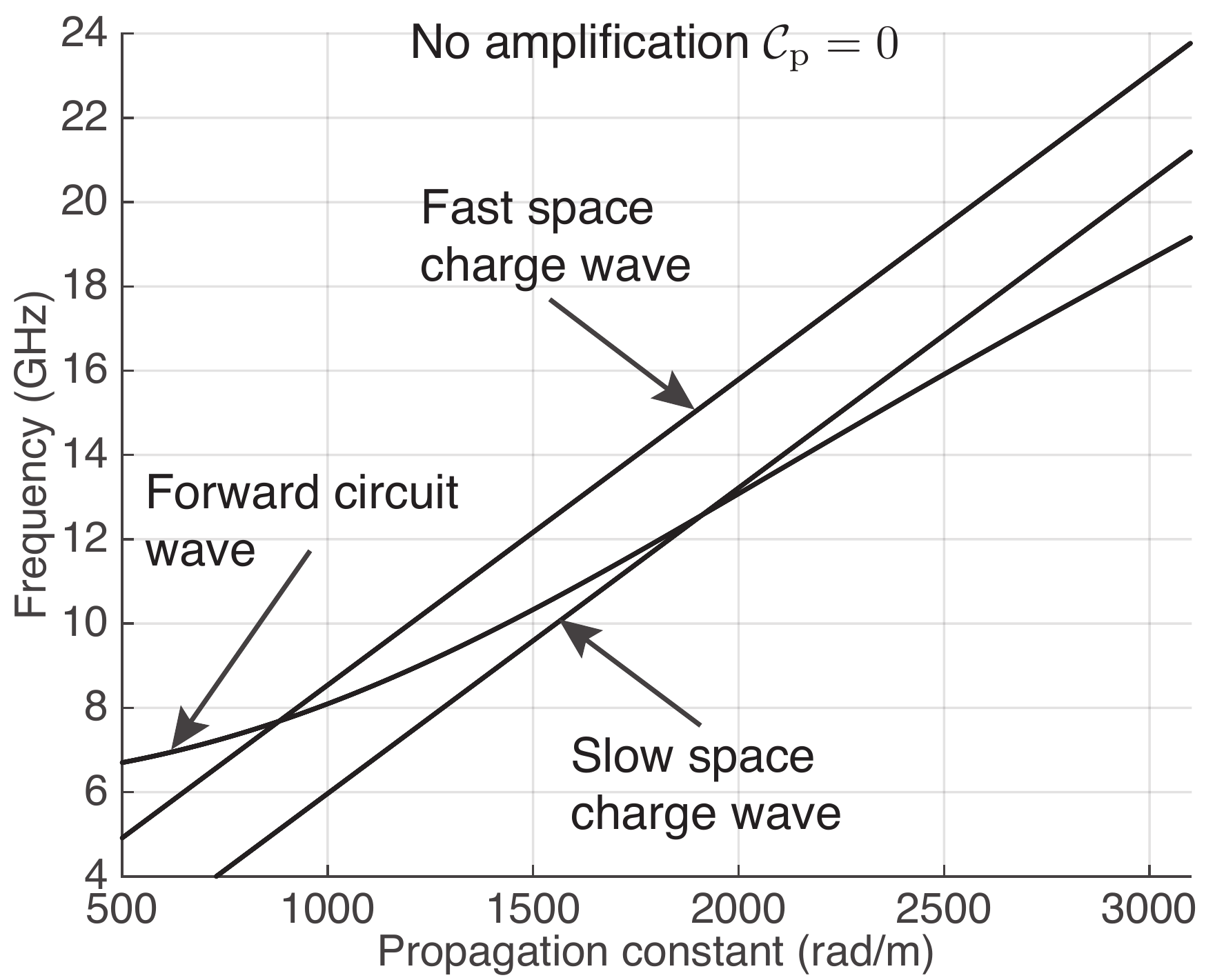}
\caption{Dispersion diagrams without coupling $\cC_{\mathrm{p}} =0$ (as if the dc current tends to zero) for the equivalent circuit model (eq.~\eqref{e:C3PierceCirq2}), or equivalently for the harmonic domain discrete model (eq.~\eqref{e:C3PierceDIMO}). Solutions are purely real values. The forward circuit wave is simply the ``cold'' dispersion relation.}
\label{f:DispRelationNoAmp}
\end{figure}

To compare accurately both models from sections \ref{s:equiCir} and \ref{s:Dimo}, we take a phase velocity $v_{\mathrm{ph},0}$ depending on our ``cold" dispersion relation, instead of taking it constant, as in Pierce's theory.

\subsection{Without amplification}

We consider the case when the Pierce parameter tends to zero
($\cC_{\mathrm{p}} \rightarrow 0$), as if the dc current also tends to
zero ($I_0 \rightarrow 0$). The four solutions of the dispersion
relations \eqref{e:C3PierceCirq2} and \eqref{e:C3PierceDIMO} of the
equivalent circuit and discrete models are identical. Solutions for
forward and return circuit waves are $\omega = \pm \beta
v_{\mathrm{ph},0}$. For the discrete model, we have $\omega = \pm
\Omega^s_{\beta}$ but because we take the same ``cold" dispersion
relation and because $\cC_{\mathrm{p}} \rightarrow 0$, we can take
$\Omega^s_{\beta} = \beta v_{\mathrm{ph},0}$, leading to identical
results for both models. Solutions for the slow and fast beam waves
are $\omega = \beta v_0 \pm \omega_\rmpp$. Those solutions are
presented in Fig.~\ref{f:DispRelationNoAmp}. 

\subsection{With amplification}

First, we notice that relations \eqref{e:C3PierceCirq2} and \eqref{e:C3PierceDIMO} coincide 
when using the first order linear approximation for numerators of the second fractions. Indeed, 
near the wave resonance (when $\omega \simeq \Omega^s_{\beta}$, 
viz.\ $\beta \simeq \beta_0$), Taylor expansion yields
\begin{align}
\omega^2  -  \beta^2 v^2_{\mathrm{ph},0} &= 2 \omega v_{\mathrm{ph},0} \left( \beta - \beta_0 \right) + \cdots \\
\omega^2 - (\Omega^s_{\beta})^2 &= 2 \omega v_{\rmg} \left( \beta - \beta_0 \right) + \cdots \, .
\end{align}
This approximation leads to the conclusion that the harmonic domain
discrete model provides the same results as the equivalent circuit
model when the dispersion diagram is a slight perturbation of the
un-coupled waves, which is the case for practical devices.

But, outside this approximation, we expect small variations between the
two models.  The maximum distance between un-coupled and coupled waves
occurs at the amplification band edges where mode coalescence takes
place. To assess them on an example, we take the ``cold'' dispersion
relation of a TWT and we solve the previous equation. 

\begin{figure}[!t]
\centering
\includegraphics[width=\columnwidth]{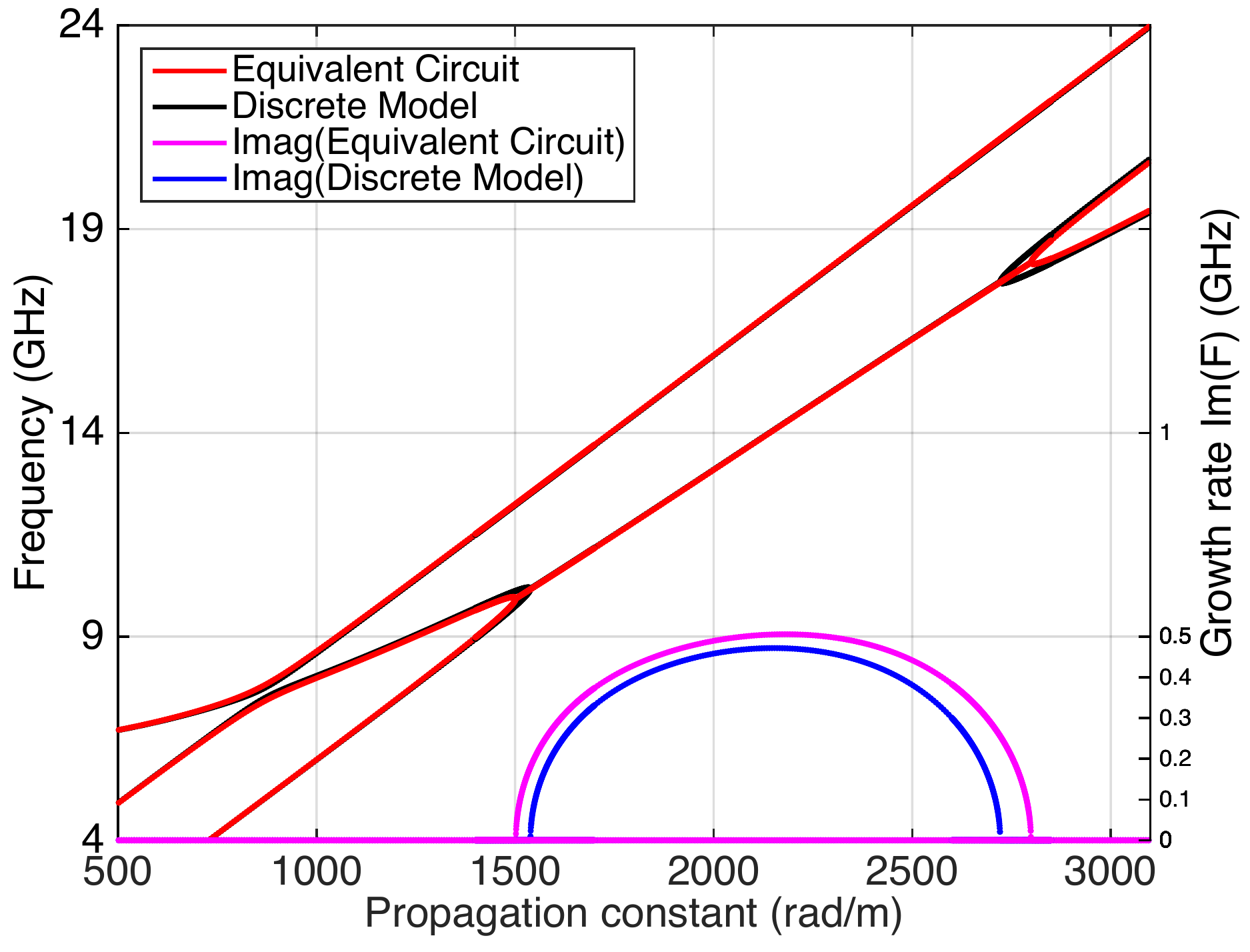}
\caption{``Hot'' linear dispersion diagrams $\cC_{\mathrm{p}} > 0$ ;
  left ordinate axis : real part of the frequency ; right axis : imaginary part. Equivalent circuit model \eqref{e:C3PierceCirq2} is in red for the real part and in magenta for the imaginary part. Harmonic domain discrete model \eqref{e:C3PierceDIMO} is in black for the real part and in blue for the imaginary part. Both models yield almost identical results, with fast space charge wave, slow space charge wave, and forward circuit wave, except at band edges. Tube passband from 10 to 18 GHz.}
\label{f:DispRelation}
\end{figure}

As the independent variable in \eqref{e:C3PierceCirq2} and \eqref{e:C3PierceDIMO} 
is the propagation constant $\beta$, amplification is considered in time, 
with complex frequencies $\omega(\beta)$ whose imaginary parts are growth rates. 
The tube passband is defined when non-zero growth rates occur. 
A symbolic solver provides solutions for the four waves as
represented with Fig.~\ref{f:DispRelation}. 
We immediately see the
close similarity between both models as their solutions are
almost superposed. The upper curve stands for the fast space charge
wave, lower curve depicts the slow space charge wave, and between them
we see the forward circuit wave. The backward circuit wave, 
with negative frequencies or negative propagation constants, is not
shown. From 10 to 18 GHz, real solutions for the slow space charge
wave and the forward circuit wave are superposed, and for both waves,
we have non-zero imaginary parts: this defines the passband of
the tube.

\begin{figure}[!t]
\centering
\includegraphics[width=\columnwidth]{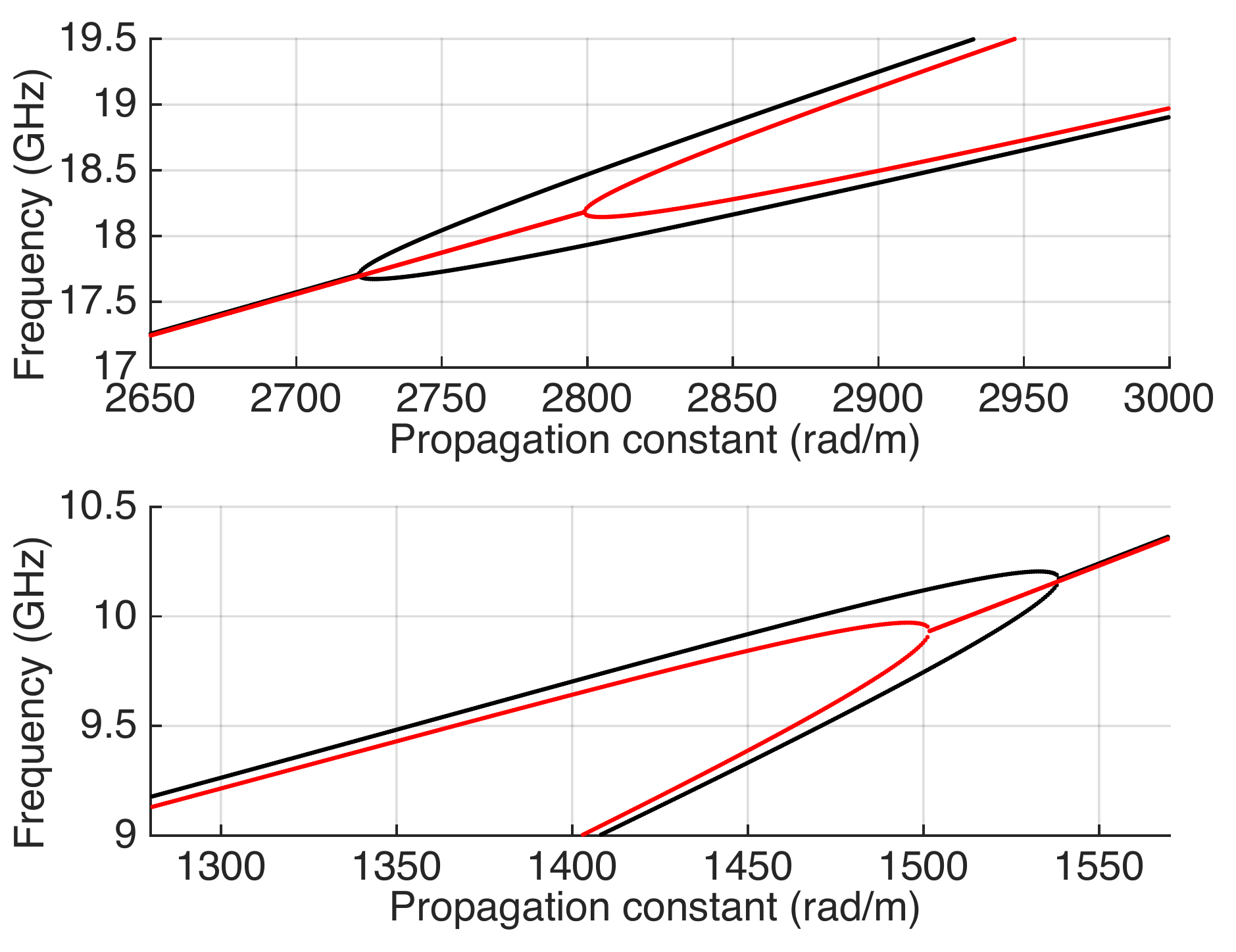}
\caption{Zoom near band edges on ``hot" linear dispersion diagrams of Fig.~\ref{f:DispRelation}. Equivalent circuit passband starts lower in frequency, near 10 GHz, when discrete model passband starts at 10.2 GHz. Equivalent circuit passband stops higher at 18.1 GHz, when the discrete model stops at 17.7 GHz.}
\label{f:DispRelation2}
\end{figure}

A zoom at band edges of Fig.~\ref{f:DispRelation} is presented in
Fig.~\ref{f:DispRelation2}. As expected, small differences occur when
we study the band edge vicinity. Similar differences, but with other
dispersion relations, were found in \cite{the16a}. The main difference
is the size of the passband: larger for the equivalent circuit.  

\section{Conclusion and perspectives}
\label{Conclusion}

We first presented another way to find the ``hot'' dispersion relation
of the Pierce equivalent circuit, using (less usual) beam and
circuit-beam impedances. After recalling the basis of the discrete model,
we computed its ``hot'' dispersion relation in linear harmonic domain.
Finally, an analytical comparison shows that both models lead to
similar results, which validates the discrete model in small signal
regime.

However, small measurable differences do exist between the models so
that one of them (or both) must deviate from the experiment. We
suggested elsewhere \cite{the16b} that the Pierce model slightly
violates Maxwell equations when coupling is strong. On the other hand, 
we see no approximation in the discrete model except the truncation on
the number of modes which is a sensible approximation. Based on these
arguments, we speculate that the Pierce model is more likely to contain
approximations than the discrete model, especially because discrepancies are stronger near the band edges, where the Pierce coupling impedance tends to infinity.
According to us, a major advantage of the discrete model is its validity near band edges as well as in the center of passband.

In the continuity of this work, frequency \cite{ter16} and time domain
\cite{min16} simulations are currently investigated in large signal
regime.

\appendix

\section{Sheath helix approximation} 
\label{s:SheaHelix}

Now, we present an application of the discrete model on the sheath helix model \cite{pie50}, a three-dimensional traveling-wave tube model. 
Starting from the real periodic structure instead of the equivalent circuit, we obtain the circuit impedance from the tube geometry.
First of all, derivation of the normalisation \eqref{e:NormalisationNsb} provides \cite{min16} the group velocity
\begin{equation} \label{e:vg}
v_{\rmg} (s,\beta) = \frac{d}{N^s_{\beta}} \int_{\cS} \Re \left(\bsE^{s*}_{\beta} (\bfr) \times \bsH^s_{\beta} (\bfr) \right) \cdot \bfe_z \upd \cS \, ,
\end{equation}
where the surface integral is equal to $1/d$ times the cell volume integral. 
In the sheath helix model, we use only one propagation mode (we omit superscrit $s=0$), without space harmonics. The flux of the Poynting vector in the harmonic discrete model along the $z$-axis reads \cite{min17a,min17b}
\begin{align} 
\langle \cP \rangle = \frac{1}{2} \Re  \frac{1}{2 \pi}  \int^{\pi}_{-\pi} \tilde{\sV}^{*}_{\beta} \rmi \tilde{\sI}_{\beta} \frac{1}{d} \sK_{\beta} c^2  \upd (\beta d)   \, ,
\end{align}
with the geometric propagation factor $\sK_{\beta}$ resulting from \eqref{e:vg}
\begin{equation}
\sK_{\beta} c^2 = v_{\rmg} \, N_{\beta}\, ,
\end{equation}
with $c$ the speed of light. Knowing the 3D boundary conditions, one
can write solutions of the Helmholtz equations \eqref{e:Helmo1} and
\eqref{e:Helmo2} for $\bsE_{\beta}(\bfr)$ and $\bsH_{\beta}(\bfr)$,
and provide a definition for each eigenfield leading to
\begin{equation}
\sK_{\beta} c^2 = \frac{\beta N_{\beta} \epsilon_0}{\gamma^2} \pi a^2 |\sE^{s}_{z,\beta}(r=0)|^2 \mathcal{F}(\gamma a) \, ,
\end{equation}
with $\pi a^2$ the disc section area of the helix, with the transverse propagation constant $\gamma = \sqrt{ \beta^2  -(\Omega_{\beta}/c)^2}$, and with the dimensionless impedance reduction factor \cite{pie47}
\begin{align}
 \mathcal{F}(\gamma a) &=  \left( 1 + \frac{I_0 K_1}{I_1 K_0} \right) \left( I^2_1 - I_0   I_2  \right) \nonumber \\
 & \quad \, + \left( \frac{I_0}{K_0} \right)^2 \left( 1 + \frac{I_1 K_0}{I_0 K_1} \right)\left( K_0 K_2 - K^2_1 \right) \, ,
\end{align}
where $I_m = I_m(\gamma a)$ and $K_m = K_m(\gamma a)$ are modified Bessel functions of the $m^{\textrm{th}}$ order of the first and second kinds respectively.
Using \eqref{e:PierceImp}, we finally recover the circuit impedance in the thick beam model as
\begin{equation} \label{e:3DZc}
Z_{\rmc}(\beta)  = \frac{1}{ \pi  a^2\epsilon_0} \frac{\gamma^2}{\Omega_{\beta} \beta^3} \Big[ \mathcal{F}(\gamma_0 a) \Big]^{-1} \, ,
\end{equation}
depending on the helix geometry \cite{pie50,tie53}. A similar development can be done for any tube geometries from the discrete model, so it is well adapted to investigating 3D structures.

\section{Beam--plasma systems} 
\label{s:BeamPlasma}

Because they generate only little noise, traveling-wave tubes have also proved to be good tools for plasma physics (beyond the fact that the beam is already a plasma). In the classic beam-plasma system \cite{one71,tsu82,tsu82prl,dov07}, waves are propagated using the classic plasma itself. To study this system, we substitute the propagating medium with a slow-wave structure like in a TWT.
Following \cite{tsu82}, considering the power definition, in harmonic domain, for a one-dimensional plasma $\langle \cP \rangle = \int v_{\rmg} \mathcal{E} \upd \cS$, where $v_{\rmg}$ is the group velocity, and $\mathcal{E}$ is the wave energy density of the plasma given by
\begin{equation}
\mathcal{E} = \frac{\epsilon_0}{2} \omega \derivep{}{\omega}\left(\varepsilon(\beta,\omega) \right)\Big|_{\omega,\beta_0}  \langle E^2_{z,\rmc} \rangle \, 
\end{equation}
from the average squared electric field, with $\varepsilon(\beta,\omega)$ the plasma dielectric function. Thus
\begin{equation}
\langle \cP \rangle = \frac{-  \pi a^2 \epsilon_0}{2}  \omega \derivep{}{\beta}  \left(\varepsilon(\beta,\omega) \right)\Big|_{\beta_0}  |E_{z,\rmc}|^2 \, ,
\end{equation}
because $v_{\rmg} = \derivep{\omega}{\beta}$.
Using \eqref{e:PierceImp}, we make the link between the beam-plasma system and the beam--slow-wave structure system by setting the plasma impedance
\begin{equation} \label{e:Zplasma}
Z_{\rmpp} = \frac{- 1}{\pi a^2 \epsilon_0} \frac{2}{\omega \beta^2} \left[ \derivep{}{\beta}  \left(\epsilon(\beta,\omega) \right)\Big|_{\beta_0} \right]^{-1}
\end{equation}
equal to $Z_{\mathrm{c}}$. Eq.~\eqref{e:Zplasma} can also be obtained from computing circuit potentials of both systems, as done in \cite{guy96}. Then the Pierce parameter becomes
\begin{equation}
\cC^3_{\mathrm{p}} = \frac{ - \omega_\rmpp^2}{ \omega \beta^2 v_0} \, \left[ \derivep{}{\beta}  \left(\epsilon(\beta,\omega) \right)\Big|_{\beta_0} \right]^{-1}  \, .
\end{equation}
We immediately see the analogy between the helix slow-wave structure circuit impedance \eqref{e:3DZc}, depending on the tube geometry, and the plasma impedance \eqref{e:Zplasma}, depending on the plasma dielectric function.
It is because waves in a TWT are expressed thank to the dispersion relation and the way they are coupled with the beam.
Ref.\ \cite{one71} takes $\beta = \beta_{\rme} = \omega / v_0$, so the linear Landau growth rate is $\gamma_{\mathrm{max}} = (n_{\rmpp}/n_{\rmb})^{1/3} \sqrt{3} \, \cC_{\mathrm{p}} \, \omega / 2^{1/3}$, with $n_{\rmb}$ and $n_{\rmpp}$ the beam and plasma densities.
Analogy of the beam-plasma and the TWT slow-wave structure is allowed by replacing the  dielectric function by the geometric factor contained in the circuit impedance.

\section*{Acknowledgement}

The authors would like to thank F.~Doveil for his suggestions.

\end{document}